\documentclass[prd,aps,twocolumn,showpacs,superscriptaddress]{revtex4}
\usepackage{amsmath}
\usepackage{amssymb}
\usepackage{amsthm}
\usepackage{mathrsfs}
\usepackage{epsfig}
\usepackage[usenames]{color}
\usepackage{graphicx}
\newcommand{\beq}{\begin{equation}}
\newcommand{\eeq}{\end{equation}}
\newcommand{\bea}{\begin{eqnarray}}
\newcommand{\eea}{\end{eqnarray}}
\newcommand{\bwt}{\begin{widetext}}
\newcommand{\ewt}{\end{widetext}}

\newcommand{\ssc}{\scriptscriptstyle}

\newcommand{\hmu}{\hat{\mu}}
\newcommand{\tr}{{\rm tr}}

\newcommand{\re}{{\rm Re}}

\newcommand{\vev}[1]{\Big\langle #1 \Big\rangle}
\newcommand{\bpsi}{\bar{\psi}}
\newcommand{\bK}{\bar{K}}

\input epsf
\begin{document}

\title{Fermion condensates and Lorentz symmetry breaking in strongly-coupled large N gauge theories}
\author{E. T. Tomboulis}
\email{tomboulis@physics.ucla.edu}
\affiliation{Department of Physics and Astronomy, UCLA,
Los Angeles, CA 90095-1547}


\begin{abstract}   
The possibility of Lorentz symmetry breaking (LSB) has attracted 
considerable attention in recent years. Spontaneous LSB, in particular, offers the attractive prospect 
of the graviton as a Nambu-Goldstone boson. 
Here we consider the question of spontaneous 
LSB in lattice gauge theories via formation of fermion condensates in the strong coupling and large N limits. 
We employ naive massless fermions in a fermionic hopping expansion in the presence of sources coupled to various condensate operators of interest. The expansion is summed in the large N limit  in two equivalent ways: (i) direct summation of 
all leading N graphs; and (ii) construction of the corresponding large N effective action for composite operators. When sources are turned off a variety of fermionic condensates is found to persist. These include the chiral symmetry breaking condensates, thus recovering previous results; but also some LSB condensates, in particular, axial vector and rank-2 tensor condensates. 
Furthermore, in the presence of internal (global) symmetry groups, formation of condensates `locking' internal and external (Lorentz subgroup) symmetries is found to also become possible. 
Some implications and open questions are briefly discussed.  
\end{abstract} 
\maketitle

\section{Introduction}
Over the last several years there has been a surge of interest in the possibility of 
explicit or spontaneous breaking of Lorentz symmetry. This has been motivated by a variety of phenomenological 
and theoretical reasons that cannot all be reviewed here. In this paper we focus on the possibility of spontaneous dynamical 
breaking of Lorentz symmetry in field theory. The idea goes back to Bjorken \cite{Bj} who proposed that the photon 
be viewed as the Goldstone boson of such breaking. The same idea was soon afterwards applied to the graviton \cite{PO}. The application to the graviton is particularly attractive. A Goldstone graviton 
offers a potential avenue to a quantum gravity theory that evades the well-known difficulties of 
quantizing the metric field of General Relativity as an elementary field. In fact the  
similarity of the basic mathematical constructs in General Relativity and  
effective Lagrangians for non-linearly realized broken symmetries was 
generally noted when chiral models were introduced in the late sixties.   

The idea of the Goldstone graviton has been revived more recently \cite{KT}, 
and a number of effective field theory analyses have been performed \cite{KT} - \cite{BK}.
In an effective field theory analysis one assumes that Lorentz symmetry breaking takes place 
at a certain scale. One then proceeds to examine the consequences at lower scales and, in particular, construct the effective theory describing the interactions of the resulting Goldstone bosons among themselves and other surviving light degrees of freedom. 
The effective theory of course involves non-renormalizable interactions.  
A central question then is whether the assumed symmetry breaking can actually take place in some  
underlying theory which is UV complete. It appears very difficult to come up with an 
UV healthy, or at least perturbatively renormalizable, model in which Lorentz symmetry 
breaking occurs  
at weak coupling. (At least this author is not aware of any such satisfactory model.) This 
may not be surprising since one naturally expects dynamical Lorentz symmetry breaking to take 
place at strong coupling. 

Here we examine this question in $SU(N)$ or $U(N)$ lattice gauge theories in the strong coupling 
and large $N$ limits. The lattice theory at strong bare coupling provides a good, tractable 
model for first exploring such non-perturbative issues. The lattice spacing represents the scale at which couplings are strong and non-perturbative dynamics takes place. 
Indeed, as is well-known, the model exhibits all the salient non-perturbative 
features of QCD-like theories, in particular confinement and chiral symmetry-breaking.   
The model is of course far from the continuum limit \cite{F0}.

Chiral symmetry breaking via formation of fermionic condensates in strongly-coupled lattice gauge theory is a prototypical example of dynamical symmetry breaking. 
It is natural then to ask whether further 
condensates can form that break other global symmetries such as Lorentz symmetry.  
 `Lorentz' symmetry here actually refers to the $SO(d)$ symmetry after Wick-rotation 
 to the $d$-dimensional Euclidean space of the lattice formulation, which is further reduced to hypercubic symmetry 
 due to the lattice discretization. This discretization, however, is irrelevant here; one is interested in 
 true dynamical breaking through condensate formation picking out particular directions in (latticized) space-time.

Formation of a variety of fermionic condensates can, in fact, be related to the structure of 
one physical quantity. 
Consider fermionic bilinears of the form $\bpsi(x)\Gamma^A\psi(x)$, where $\Gamma^A$ stands 
for any Clifford algebra element. The vacuum expectation value (vev) of such a bilinear is related to the fermion 2-point function $G(x,y) = \vev{\psi(x)\bpsi(y)}$ at 
$x=y$:
\beq 
\vev{\bpsi(x)\Gamma^A \psi(x)}  =  -\tr \left[ G(x,x) \Gamma^A\right]  \,,  \label{exp0}
\eeq
where the trace is over spinor and color indices. Thus, a condensate for 
$\bpsi(x)\Gamma^A\psi(x)$ will form if $\tr_{\bf \ssc C} G(x,x)$ 
has non-vanishing projection onto $\Gamma^A$. Note that $\tr_{\bf \ssc C} G(x,x)$, where  $\tr_{\bf \ssc C}$ denotes trace over color, is a gauge invariant quantity. 
Determining the structure of $G(x,x)$ then allows one to examine the formation of 
various condensates, e.g. the chiral condensate for scalar $\Gamma^A={\bf 1}$, or Lorentz breaking 
condensates for, say, vector or axial vector $\Gamma^A$.  
Condensates involving nearest neighbor lattice sites (derivatives in the continuum) can also, as 
we will see, be related to $G(x,x)$. 

In this paper we derive an equation for determining $G(x,x)$ in the strong coupling and 
large $N$ limits (section \ref{section2}). The large $N$ limit allows one to identify a well-defined infinite set of dominant graphs in a fermion hopping expansion that can then be summed. The summation may be 
performed by diagrammatic means (cf. \cite{BBEG}, \cite{MS}) leading to a self-consistent equation for the full $G(x,x)$ in this limit (subsection II A). Alternatively, and more elegantly, the same equation follows from 
directly constructing and then varying the effective action for composite operators \cite{CJT} at large 
$N$ (subsection II B).  
From this equation one recovers previous results on the formation of chiral symmetry breaking 
condensates \cite{BBEG} - \cite{KS}. One finds, however, that other condensates may also form 
via (\ref{exp0}), in particular, axial vector and tensor Lorentz-breaking condensates (section \ref{T}). 

When internal (global) symmetry groups are present another possibility arises, namely, `locking' of internal and external groups via condensate formation (section \ref{L}). Locking among internal 
groups is common. Chiral symmetry breaking in QCD is in fact an example: the left and right 
chiral rotation groups  are locked by the chiral condensate to equal rotations forming the 
unbroken diagonal subgroup. Color superconductivity arising from the formation of condensates  
locking color and flavor symmetries \cite{ARW} provides a more elaborate example. 
Locking of internal and external groups is not normally considered. The one well-known example is the locking of angular momentum and isospin in the field of a magnetic monopole \cite{JR}. 
Internal-external-group locking condensates can serve as a dynamically generated vierbein field.  
This presents an approach to a quantum theory of composite (Goldstone) gravitons that has not been explored before. 
In the context of the strongly coupled gauge models considered here 
internal-external locking presents no special problems in Euclidean space where groups are compact. Upon rotation to Minkowski space, however, where external groups such as the Lorentz group become decompactified, obvious problems can arise, in particular with respect to unitarity. This is discussed in section \ref{L}. The concluding section \ref{D} provides some further discussion of these results,  
open questions and future directions. A condensed account has appeared in \cite{T}.

\vspace{2cm}

\section{Large $N$ summation and effective action  \label{section2}}

We use standard lattice gauge theory notations and conventions. We work on a euclidean 
hypercubic d-dimensional lattice 
with lattice sites denoted by their lattice coordinates $x=(x^\mu)$,  and lattice unit vectors in the $\mu$-th direction by $\hmu$. We generally indicate dimension dependence by $d$ even though we 
are actually interested only in the $d=4$ case. 
The gauge field bond variable $U_b$ on bond $b=(x,\hmu)$ is more explicitly denoted by $U_\mu(x)$, and the fermion fields on site $x$ by $\bpsi(x)$ and $\psi(x)$. The gauge group is taken to be $SU(N)$ or $U(N)$, but the method developed below (subsections 2.1 - 2.2) can be applied to other groups. The euclidean gamma matrices satisfying
$\{\gamma^\mu, \gamma^\nu\} = 2 g^{\mu\nu}{\bf 1}$ with  $g^{\mu\nu}=\delta^{\mu\nu}$ 
are hermitian, $\gamma^{\mu\,\dagger} = \gamma^\mu$. We also define ($d=4$): $\gamma^5=
\gamma^1\gamma^2\gamma^3\gamma^4$, so that $\gamma^{5\,\dagger} = \gamma^5$.

We employ the lattice action with naive massless fermions:
\bwt
\beq
S= \sum_p \beta\,[1- {1\over N}\re \,\tr U_p ] + \sum_{b=(x,\mu)}{1\over 2} \,\left[ \bar{\psi}(x) \gamma_\mu U_\mu(x) \psi(x+\hat{\mu}) - \bar{\psi}(x+\hat{\mu}) \gamma_\mu U^\dagger_\mu(x) \psi(x) \right]  \,.
\label{act1}
\eeq
\ewt
Naive fermions, which automatically provide an anomaly-free chirally invariant model, are indeed 
well-suited for our purposes since fermion doubling is irrelevant here - in fact, as it turns out, the 
more degrees of freedom (color and flavor) the better. The use of naive versus other fermion 
formulations is further discussed in section \ref{T}. 
We will be concerned with expectations of operators of the form $\bpsi(x)\Gamma^A\psi(x)$, where 
$\Gamma^A$ may stand for a Clifford algebra element, such as 
$\Gamma_S=1$, $\Gamma^\mu_V=\gamma^\mu$, or
 $\Gamma^\mu_{A}=i\gamma^5\gamma^\mu$, or some other choice. Operators involving nearest neighbor sites (derivatives in the continuum) will also be considered below.  
It should be noted that any such Lorentz-breaking condensates may also violate some discrete symmetries. Thus, for example, a non-vanishing vector condensate would also violate $C$, whereas an axial vector condensate would violate $P$.

Since the operator $\bpsi(x)\Gamma^A\psi(x)$ is a fermion bilinear its vev is related to the 
fermion 2-point function (full propagator) $G^{a,b}_{\alpha ,\beta }(x,y)=\vev{\psi^a_\alpha (x)
\bpsi^b_\beta(y)}$ in the limit $x=y$:
\bea
\vev{\bpsi(x)\Gamma^A \psi(x)} & = & -\tr \left[ G(x,x) \Gamma^A\right] \label{exp1}
\\
           & = & -\tr_{\ssc D} \left[ \bar{G}(x,x) \Gamma^A\right] \label{exp1a}
\eea
with the second equality written explicitly in terms of the gauge invariant quantity 
$\bar{G} (x,x)\equiv \tr_{\ssc C} G(x,x)$. Here 
$\tr$ denotes trace over spinor and color indices, whereas $\tr_{\ssc C}$ and $\tr_{\ssc D}$ 
denote traces over color (Latin letters) and Dirac spinor (Greek letters) indices, respectively.

To study such expectations we add an external source $K^A$ coupled to the operator $\bpsi(x)\Gamma^A\psi(x)$. More generally, one introduces a source for $G(x,x)$ of the form $K=\bK {\bf 1_{\ssc C}}$, where ${\bf 1_{\ssc C}}$ denotes the unit matrix in color space, and $\bK$ is an arbitrary (invertible) matrix in spinor space. Coupling to any one particular fermion bilinear then amounts to a 
particular form of $\bK$; for example, $\bK = k n^\mu \gamma_\mu$, where $k$ is an arbitrary number and $n^\mu$ an arbitrary unit vector, gives a source of magnitude $k$ and direction $n^\mu$ coupled to $\bpsi(x)\gamma_\mu\psi$.

We write the action (\ref{act1}) in the presence of the external source more concisely in the 
form 
\beq 
S= \sum_p \beta\,[1- {1\over N}\re \,\tr U_p ] + \sum_{x,y} \bpsi(x){\cal K}_{x,y}(U)\psi(y) \; ,\label{act2} 
\eeq
where 
\beq 
{\cal K}_{x,y}(U) = {\bf M}_{x,y}(U) + {\bf K}_{x,y}  \label{act3}
\eeq
with
\bea
{\bf M}_{x,y}(U) & \equiv & \!\!{1\over 2}\!\!\left[ \gamma_\mu U_\mu(x) \delta_{y,x+\hmu} 
\!-\!
\gamma_\mu U^\dagger_\mu (x-\hmu) \delta_{y,x-\hmu} \right] \label{act3a}\\
{\bf K}_{x,y} & \equiv & K\, \delta_{x,y}= \bK {\bf 1_{\ssc C}} \,\delta_{x,y} \;. \label{act3b}
\eea
Note that ${\bf K}$ and ${\bf M}$ are matrices in spinor and color space as well as in lattice 
coordinate space. If the fermions 
in (\ref{act1}) are taken to also carry a flavor index,  both (\ref{act3a}) and (\ref{act3b}) should be multiplied by the flavor unit  matrix ${\bf 1_{\ssc F}}$.

In the strong coupling limit $\beta\to 0$ the plaquette term in (\ref{act2}) is dropped. The 
corrections due to this term can be taken systematically into account within the strong coupling cluster expansion, which, for sufficiently small $\beta$, converges. Hence they do not produce any qualitative change in the behavior obtained below at $\beta\to 0$. 

Setting $\beta=0$ in (\ref{act2}) then, $G(x,x)$  is given by 
\bwt
\bea
G(x,x) & = & {1\over \int [DU] \, {\rm Det} {\cal K}(U)}
\int [DU] \,  {\rm Det}{\cal K}(U)\,
{\cal K}^{-1}_{x,x}(U)
\label{exp2a}\\
& = & {
\int [DU] \,  {\rm Det} [{\bf 1 + K^{-1} M}(U)]  \; 
   \left[[{\bf 1 + K^{-1} M}(U)]^{-1} {\bf K}^{-1}\right]_{x,x}
\over \int [DU] \, {\rm Det} [{\bf 1 + K^{-1} M}(U)] }  \,,    \label{exp2b}
\eea
\ewt
in the presence of arbitrary source $K$. 
The vev of $\bpsi(x)\Gamma^A\psi(x)$ in the presence of the source is then 
obtained from (\ref{exp1}). The part projected out in (\ref{exp1}) can be picked out at the outset 
by restricting the source to the appropriate form coupling to the operator of interest.

\subsection{Large $N$ graph summation}

 We evaluate (\ref{exp2b}) in the hopping expansion. This amounts to expanding (\ref{exp2b}) treating 
 ${\bf M}$ as the interaction and taking ${\bf K}$ as defining the inverse bare propagator:
 \beq
 {\bf K}^{-1}_{x,y} = \bK^{-1}{\bf 1_{\ssc C}}\, \delta_{x,y} \,.
 \label{bareG}
 \eeq

 The textbook version of the expansion is the scalar case where ${\bf K}$ is a mass term.  
 Note that  ${\bf K}$ is purely local, whereas ${\bf M}$ has only 
nearest-neighbor non-vanishing elements ${\bf M}_{x, x+\hmu} =  {1\over 2}\gamma_\mu U_\mu(x) $ and 
${\bf M}_{x, x-\hmu} = -{1\over 2}\gamma_\mu U^\dagger_\mu(x-\hmu)$.  
In the absence of the plaquette term, and since ${\bf M}$ is linear in the bond variables $U_b$, integration over the gauge field results in non-vanishing contributions only if at least two $M$ factors with 
equal (mod $N$) number of $U_b$ and $U_b^\dagger$'s 
occur on each bond $b$. 
\begin{figure}[ht]
\includegraphics[width=\columnwidth]{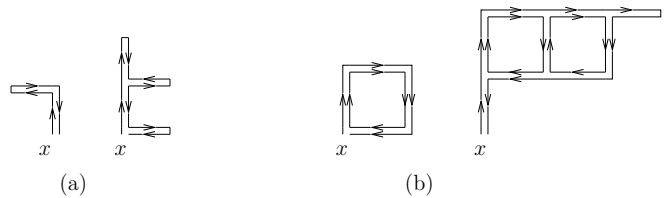}
\caption{(a) Some tree graphs; and (b) some loop graphs attached to site $x$.  \label{lsb6a}}
\end{figure}

The  expansion of the ${\cal K}^{-1}_{x,x}(U)$ is represented by all paths starting and ending at $x$, whereas that of 
the ${\rm Det} {\cal K}(U)$ by all closed paths \cite{R}, \cite{S}.  Consistent with the above constraint on each bond resulting from the $U$-integrations,  after the cancellation of all disconnected graphs between numerator and denominator the connected graphs giving the expectation 
(\ref{exp2b}) naturally fall into two classes: `tree graphs' and `loop graphs'. 

The tree graphs consist of paths starting and ending at $x$ and 
enclosing zero area (Fig. \ref{lsb6a} (a)); they arise entirely from the expansion of the ${\cal K}^{-1}_{x,x}(U)$. 
The loop graphs, such as those depicted in Fig. \ref{lsb6a}(b), consist of paths from the expansion of ${\cal K}^{-1}_{x,x}(U)$ and loops from that of ${\rm Det} {\cal K}(U)$ coupled by the $U$-integrations in the 
numerator in (\ref{exp2b}) \cite{F1}. 
We note in passing the well-known fact (see e.g. \cite{R}) concerning the hopping expansion that there are no restrictions on how many times a bond is revisited in drawing all such possible 
connected graphs \cite{F2}.

Now, the set of tree graphs are the leading contribution in $N$. Loop graphs are down by powers 
of $1/N$ relative to tree graphs \cite{BBEG}. Thus, the set of tree graphs in the hopping  
expansion give the 
large $N$ limit of the theory. The sum of all tree graphs attached at site $x$ then 
constitute the full propagator $G(x,x)$ in this limit.

The lowest order contribution is just the bare propagator ({\ref{bareG}). The full 
set of trees at $x$ is generated by the expansion of 
$ \left[[{\bf 1 + K^{-1} M}(U)]^{-1} {\bf K}^{-1}\right]_{x,x}$ 
in (\ref{exp2b}). 
We first consider trees extending from site $x$ to nearest-neighbor (nn) sites. 
The simplest such tree has only one `trunk' extending 
to any one of the $2d$ nn sites to $x$ (Fig. \ref{lsb5a}) and gives: 
\bea
G_{nn}^{(1)}(x,x) =   \sum_{\pm\hmu} \int 
dU_{x,x+\hmu} \,  \qquad \qquad\qquad & &      \nonumber \\
\cdot  \left[\,{\bf K}_{x,x}^{-1}{\bf M}_{x,x+\hmu}(U){\bf K}^{-1}_{x+\hmu,x+\hmu}{\bf M}_{x+\hmu,x}(U){\bf K}^{-1}_{x,x}\right] .  & & \label{Tnn1}
\eea
\begin{figure}[ht] 
\includegraphics[width=\columnwidth]{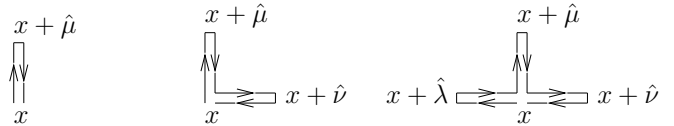}
\caption{Nearest-neighbor trees with one, two or three `trunks' at site $x$.  \label{lsb5a}}
\end{figure}

But there are also contributions from nn trees with $n$ trunks (Fig. \ref{lsb5a}), with each trunk extending to any one of the $2d$ nn sites, given by 
\bwt
\beq
G_{nn}^{(n)}(x,x) =  \sum_{T_{nn}^{(n)}} \int \prod_{b\in T_{nn}^{(n)}}
dU_b \, \left[\,\left[\prod_{j=1}^n {\bf K}_{x,x}^{-1}{\bf M}_{x,x+\hmu_j}(U){\bf K}^{-1}_{x+\hmu_j,x+\hmu_j}{\bf M}_{x+\hmu_j,x}(U)\right]\, {\bf K}^{-1}_{x,x}\right].
\label{Tnnn}
\eeq
\ewt
The sum is over the set $T_{nn}^{(n)}$ of such $n$-trunk nn trees obtained by letting each trunk 
independently extend in all possible $\pm\hmu$ directions from site $x$ to a nn site.

The $U$-integrations in (\ref{Tnn1}) and (\ref{Tnnn}) are essentially trivial since, as it is easily seen, 
the product of $U_b$'s along the closed path forming each tree yields the unit matrix in 
color space. (This is true for any, not just nearest-neighbor,  trees.) 
\begin{figure}[ht]
\includegraphics[width=0.75\columnwidth]{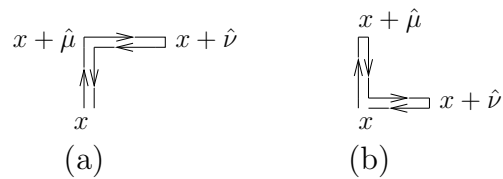} 
\caption{The nearest-neighbor 2-trunk tree at site $x$ in (b) is contained in the set of 1-trunk trees 
at $x$ depicted in (a) if all $2d$ directions are allowed at $x+\hmu$, i.e. backtracking is allowed.   \label{lsb3}}
\end{figure}

Starting from these nearest-neighbor trees the set of all trees at $x$ can be generated in a recursive manner \cite{BBEG} by attaching trees at each site 
$x+\hmu_j$, $(j=1,\ldots, n)$ of every 
n-trunk nn-tree at $x$. 
As noted in \cite{MS}, however, this grouping does not give a one to one labeling of 
the set of all trees unless a further constraint is introduced. This is illustrated in Fig. \ref{lsb3}. 
\begin{figure}[ht]
\includegraphics[width=0.35\columnwidth]{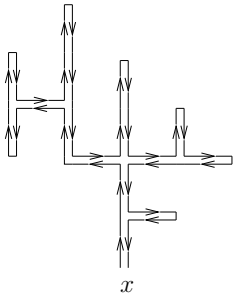}  
\caption{A general irreducible tree (IT) at site $x$: at each branching only further IT graphs can be attached, i.e. at each branch junction no backtracking along the preceding step is allowed leaving $(2d-1)$ available directions to move in.   \label{lsb4}}
\end{figure}

To obtain a one to one labeling one proceeds as follows. First note that any tree graph 
attached at $x$ is specified by a sequence of directions $\hmu_1, \hmu_2, \hmu_3, \ldots, \hmu_{2l}$  tracing a path of $2l$ steps starting and ending at $x$ as, e.g., in Fig \ref{lsb6a} (a). 
An irreducible tree (IT) graph at $x$ is now defined \cite{MS} as a tree for which this sequence cannot be truncated 
at some intermediate step $\hmu_k$ and still result in a tree graph attached at $x$. 
A general IT graph attached at $x$ is then specified by the initial direction $\hmu_1$ followed by 
attaching a sequence of IT graphs at $x+\hmu_1$; the tree is then completed by the last step necessarily in the direction $-\hmu_1$ from $x+\hmu_1$ back to $x$. 
This is illustrated in Fig. \ref{lsb4}.

If the sum of all IT graphs at a point $y$ is denoted by $G_I(y,y)$, this 
recursive building of irreducible trees immediately implies the self-consistency relation graphically represented in Fig. \ref{lsb5}. 
Using the explicit expressions (\ref{bareG}), (\ref{act3a}) the graphical equation in Fig. \ref{lsb5}     gives:   
\bea
G_I & = & \left[ {\bf 1_{\ssc C}} + \sum_{n=1}^\infty \left[ {(-1)\over 4} \bK^{-1} 
{(2d - 1) \over 2d} 2\gamma^\mu 
G_I \gamma_\mu \right]^n  \right] \bK^{-1}  
\nonumber \\
    & = & \left[ {\bf 1_{\ssc C}}  +   {1\over 2}\bK^{-1} {(2d - 1) \over 2d} \gamma^\mu G_I \gamma_\mu 
 \right]^{-1} \bK^{-1} \nonumber \\
    & = & \left[ \bK {\bf 1_{\ssc C}} + {1\over 2}\, {(2d - 1) \over 2d}\gamma^\mu G_I \gamma_\mu \right]^{-1} \,.
 \label{GIfull} 
\eea 
Note that, for space-independent source $K$, $G_I(x,x)$ is in fact $x$-independent by translation invariance.  
The hopping expansion which, in the large $N$ limit, gave the series in the first equality in 
(\ref{GIfull}) 
converges for sufficiently large $||K||$. The summed expression (\ref{GIfull}), however, can be continued to all $K$. In particular, one is interested in possible solutions to (\ref{GIfull}) 
for $K\to 0$.

Let $G^{(n)}(x,x)$ denote the `full n-bottom-trunk' tree  defined by attaching the complete set of IT graphs, i.e. $G_I$, at each site $x+\hmu_j$, $j=1,2,\ldots,n$, of every n-trunk nn tree at $x$. 
This amounts to replacing ${\bf K}^{-1}_{x+\hmu_j,x+\hmu_j}$ in (\ref{Tnnn}) by $G_I(x+\hmu_j,x+\hmu_j)$. 
The complete set of trees at $x$, comprising the full tree propagator $G(x,x)$, is now 
recovered by summing all these 'full n-bottom-trunk' trees  including the zeroth-order $n=0$ 
(no bottom trunk, i.e. bare propagator (\ref{bareG})) term: 
\beq
G(x,x) = {\bf K}_{x,x}^{-1} + \sum_{n=1}^\infty \,G^{(n)}(x,x) \, . \label{TG}
\eeq
This is represented graphically in Fig. \ref{lsb9d}. Explicitly, (\ref{TG}) then gives:
\bea
G & = & \left[ {\bf 1_{\ssc C}} +  \sum_{n=1}^\infty \left[ {(-1)\over 4} \bK^{-1} 
2\gamma^\mu 
G_I \gamma_\mu \right]^n  \right] \bK^{-1}  
\nonumber \\
    & = & \left[ {\bf 1_{\ssc C}}  +   {1\over 2}\bK^{-1} \gamma^\mu G_I \gamma_\mu 
 \right]^{-1} \bK^{-1} \nonumber \\
    & = & \left[ \bK {\bf 1_{\ssc C}} + {1\over 2}\, \gamma^\mu G_I \gamma_\mu \right]^{-1} \,.
 \label{Gfull} 
 \eea
Note that, since all trees, and hence the sum $G_I$, are diagonal in color space (cf remarks following (\ref{Tnnn})) equations (\ref{GIfull}), (\ref{TG}) are essentially equations for the gauge invariant quantity $\bar{G} \equiv \tr_{\ssc C} G$.

It is of some interest to note how the result (\ref{GIfull}), (\ref{Gfull}) for this large $N$ summation is altered were one to ignore the labeling ambiguities dealt with by 
the introduction of ITs. This is in fact what was done in the original argument given in \cite{BBEG}. It simply amounts to attaching all trees rather than only ITs  at each nn tree, 
resulting in a slight overcounting of graphs, and (\ref{Gfull}) being replaced by 
\bea
G & = & \left[ {\bf 1_{\ssc C}} +  \sum_{n=1}^\infty \left[ {(-1)\over 4} \bK^{-1} 
2\gamma^\mu G \gamma_\mu \right]^n  \right] \bK^{-1}  
\nonumber \\
    & = & \left[ \bK {\bf 1_{\ssc C}} + {1\over 2}\, \gamma^\mu G \gamma_\mu \right]^{-1} \, , 
 \label{Gfull-naive} 
 \eea
i.e. a self-consistent equation directly for $G$. It is clear that if non-trivial solutions to (\ref{GIfull}) exist for a particular condensate operator, so do they in the case of (\ref{Gfull-naive}), and vice-versa. Indeed the predictions of the exact equations (\ref{GIfull}), (\ref{Gfull}) for the various condensates considered below differ from those of (\ref{Gfull-naive}) only by 
inessential numerical factors (see section 3). These are the correction factors due to the small volume exclusion effect included in the ITs and not taken into account in (\ref{Gfull-naive}).

\subsection{Large N composite operator effective action} 

An alternative method to summing the (infinite) set of graphs 
contributing to the expectation in the large $N$ limit is the direct construction of the 
corresponding effective action. By the latter we mean the standard field-theoretic definition of 
effective action, i.e. that functional of the 
expectation of an operator which is  defined as the Legendre transform of the free energy with respect to the source coupled to the operator. Since here we deal with composite, viz. bilinear fermion operators,  
this is the effective action for composite operators \cite{CJT}. 
\bwt
\ 
\begin{figure}[ht]
\includegraphics[width=0.75\columnwidth]{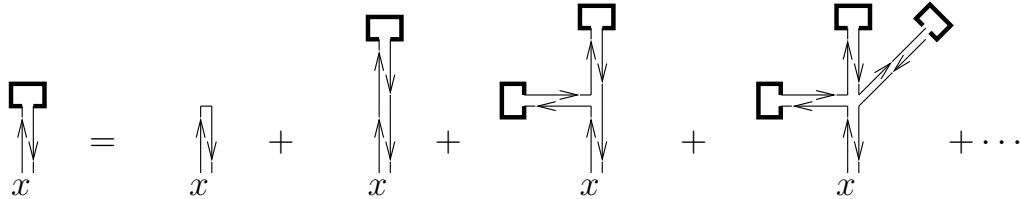} 
\caption{Self-consistent equation for the sum $G_I$ (heavy square) of ITs.  The first branch 
extends from $x$ to the nearest neighbor site in some direction $\hmu$. On the r.h.s., for each subsequent branch attached at $x+\hmu$ no backtracking along the direction of the previous step is allowed and summation over the allowed $(2d-1)$ directions is understood.  \label{lsb5}}
\end{figure}

\ 
\begin{figure}[ht] 
\includegraphics[width=0.55\columnwidth]{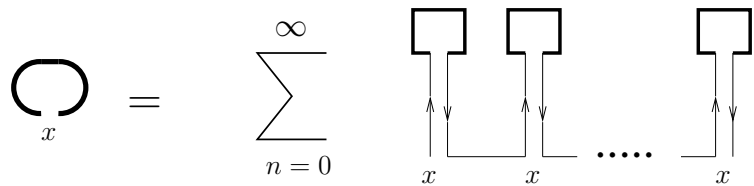} 
\caption{The equation for the sum of trees attached at $x$, i.e. $G(x,x)$ (heavy circle).  On the r.h.s., even though drawn separated for convenience, 
all trunks are connected locally in sequence at site $x$, and independent summation of each trunk over the available $2d$ directions is understood.  \label{lsb9d}} 
\end{figure}
\ewt

We note in passing that the effective action can be constructed graphically by proceeding further along the lines of the previous subsection. 
The tree graph summation there was effected by replacing the bare propagator by a dressed propagator on top of each trunk in the general nn tree. 
Complete replacement of all bare propagators by dressed propagators in the set of graphs for an expectation, or, more appropriately, the vacuum graphs giving the free energy,  
is not straightforward as double-counting has to be taken properly into account.  
But it can be carried out systematically to construct the effective action.  A much more convenient and concise derivation, however, is obtained by functional techniques, and the final result is easily  stated \cite{CJT}. 

We can straightforwardly apply the general recipe for the effective action given in 
\cite{CJT} to the theory (\ref{act1}) in the strong coupling limit to obtain the effective action 
for our object of interest $G(x,x)=\vev{\psi(x)\bpsi(x)}$. We write the action 
(\ref{act1}) without the plaquette term with the addition of a source $K=\bK {\bf 1_{\ssc C}}$:  
\beq
S= \sum_{x,y} \bpsi(x){\bf M}_{x,y}(U)\psi(y) + \sum_x \bpsi(x) K\psi(x)  \, ,\label{act4} 
\eeq
with ${\bf M}_{x,y}$ defined in (\ref{act3a}). 
The Legendre transform of the free energy 
\beq
W[K] = -\ln Z[K]= - \ln  \int  [DU] [D\bpsi][D\psi] \,e^{-S(U,\bpsi,\psi)} \label{fe}
\eeq
defines the effective action:  
\beq
\hat{\Gamma}[G]=W[K] + \sum_x \tr G(x,x)K \, .\label{Ltransf}
\eeq
Apart from the source term everything in (\ref{act4}) is treated as interaction terms, which in fact makes for a simplified version of the general expression in \cite{CJT}. 
Apart from an inessential additive constant then, $\hat{\Gamma}[G]$ is given by 
\beq
\hat{\Gamma}[G] = \sum_x \tr \ln G(x,x)- \hat{\Gamma}_2[G]   \, , \label{EA1}
\eeq
where $\hat{\Gamma}_2[G]$ denotes the sum of all 2-particle-irreducible vacuum graphs 
computed with the interactions defined in (\ref{act4}). 
In classifying these graphs, however, let us, for simplicity, ignore the slight overcounting 
resulting from dropping any restrictions on backtracking in trees. 
Then the large $N$ limit results into a great simplification: only one graph 
(Fig.\ref{lsb11a})) contributes to $\hat{\Gamma}_2[G]$. 
\begin{figure}[ht] 
\includegraphics[width=0.17\columnwidth]{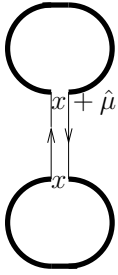} 
\caption{Single graph in terms of $G$ contributing to the effective action at large $N$ \label{lsb11a}} 
\end{figure}
Its straightforward evaluation gives then 
\beq
\hat{\Gamma}_2[G] = {1\over 4}\sum_{x,\mu} \tr \, G(x,x)\gamma^\mu G(x+\hmu, x+\hmu) \gamma_\mu
 \, \label{EA2}
\eeq 
Seeking  translation-invariant solutions to the variational equation
\beq
{\delta \over \delta G(x,x)} \hat{\Gamma}[G] =  K \, ,
\eeq
now gives 
\beq
G^{-1} = \left[ \, K + {1\over 2}   \gamma^\mu G \gamma_\mu 
\,\right]  \, . \label{EA3}
\eeq
This reproduces (\ref{Gfull-naive}). As already remarked, and will be seen explicitly in the following section, use of (\ref{EA3}) gives, except for some small numerical corrections, the same results as the exact treatment leading to (\ref{Gfull}). Identical results to those of the direct summation of the previous section are thus obtained, as 
expected. This is in fact a nice example of the elegance and efficacy of the effective action formalism:  only one graph needs to be considered instead of the infinite summation of the 
previous section \cite{F4}.

\subsection{Use of alternate approaches} 
Our approach above was based on (\ref{exp2a}) - (\ref{exp2b}) which was obtained by integrating out the fermion fields.  Next, carrying out the gauge field integrations within the hopping expansion, the set of dominant graphs in the large $N$ limit could be identified and summed either directly or by construction of the corresponding composite operator effective action. 
Here we want to briefly comment on a related alternative approach that has also been used 
to discuss chiral symmetry breaking in the same limit \cite{KS}. It is based on the fact that in the strong coupling limit the action becomes a sum of one-bond terms allowing one, in principle, to first carry out the integrations over each gauge field bond variable: 
\bea
Z[K] & = & \int  [DU] [D\bpsi][D\psi] \,e^{-S(U,\bpsi,\psi)}  \nonumber \\
& = & \int [D\bpsi][D\psi] \,e^{[w(\bar{A}, A) 
- \bpsi K\psi]}    \,, \label{SD1}
\eea 
where $A_{\mu b}^a\equiv {1\over 2}\bar{\psi}_b(x+\hat{\mu}) \gamma_\mu \psi(x)^a$ and  
$\bar{A}_{\mu b}^a\equiv -{1\over 2}\bar{\psi}_b(x) \gamma_\mu \psi(x+\hat{\mu})^a$. 
Expressions for $w(\bar{A}, A)$ are known in several cases \cite{ESS} and, in particular, 
for $U(N)$ in the $N\to \infty$ limit \cite{BG}. In the latter case, the so-called strong-coupling phase expression for $w(\bar{A}, A)$ is given in terms of functions that can be expanded in 
series in powers $\tr (\bar{A}A)^k$, $k>0$, \cite{BG}. 
The method in \cite{KS} consists of introducing 
\beq 
N{\cal M}(x) = \tr_{\ssc C}\psi(x) \bpsi(x) \, , \label{mes}
\eeq
re-expressing powers of  $\tr [\bar{A}A/N^2]^k$ in terms of ${\cal M}$ 
 and then rewriting (\ref{SD1}) in the form 
\bea
Z[K] 
&= &  e^{w(\frac{1}{N}{\partial \over\partial K})} \int [D\bpsi][D\psi] \,e^{ \tr [(\psi\bpsi)K]}  
\nonumber \\ 
& = & e^{w({1\over N}{\partial \over\partial K})} 
\int [D{\cal M}] \,e^{ [w_0({\cal M}) + N\tr{\cal M} K]}   \label{SD2} \\
&=& \int [D{\cal M}] \,e^{\left[w({\cal M}) + w_0({\cal M}) + N\tr{\cal M} K \right]} \, . 
\label{SD3}
\eea
In (\ref{SD2}) the free (pure source) fermionic integral is first formally rewritten as a bosonic 
integral in terms of ${\cal M}$ with $w_0({\cal M})= [- N\tr\ln{\cal M} + {\rm constant}]$, leading to the full 'bosonized' form (\ref{SD3}) \cite{F5}. 
The final step is to argue that in the 
$N\to \infty$ limit it suffices to evaluate the integral (\ref{SD3}) in the saddle-point 
approximation, i.e. one determines the expectation of ${\cal M}$ as the stationary points of the action 
in the integrand in (\ref{SD3}).  

The method yields essentially the same results (with the provisos of 
footnote \cite{F5}) as above. It is, however, restricted 
for $N\to \infty$ to just $U(N)$. Furthermore,  due to the unwieldy 
form of $w$ the ensuing computations are ugly and rather non-illuminating compared to the 
cleaner direct schemes above, and will not be considered here any further.

\section{Condensate formation}\label{T}

An operator whose expectation can be related to $G(x,x)$, as, for example, in (\ref{exp1}), 
may acquire a non-vanishing vev through a non-trivial solution to (\ref{GIfull}). This will happen if 
the expression relating the vev to $G(x,x)$ projects out a non-vanishing part of the solution.

We will not make a general study of solutions to (\ref{GIfull}) here. It suffices to 
consider particular branches of solutions that are picked out by appropriate 
choice of the source $\bK$. 
In all cases, at large $||\bK||$ a solution always exists that reproduces 
the hopping expansion perturbative solution. This is clear from the construction that lead to 
(\ref{GIfull}), (\ref{Gfull}). We are, however, ultimately interested in solutions at vanishing external sources.

In the scalar case, $\bK=k{\bf 1_{\ssc D}}$, the solution is $G_I=g_I(k){\bf 1_{\ssc D}}{\bf 1_{\ssc C}}$, with, from (\ref{GIfull}),  
$g_I(k)$ satisfying 
\beq
{(2d-1)\over 4} \, g_I^2 +kg_I -1 =0 \,.\label{g1}
\eeq
Then, from (\ref{Gfull}), $G=g(k) 1_{\ssc D} {\bf 1_{\ssc C}}$ with 
$g(k) = [
k + d \,g_I(k)/2]^{-1} $. 
Solving (\ref{g1}) one finds $g(0)=\sqrt{(2d-1)}/d$, and hence a scalar condensate \cite{F6} 
\beq
\vev{\bpsi(x)\psi(x)} = -NS \sqrt{{2\over d}}\, \sqrt{1-{1\over 2d}}\; , \label{Scond}
\eeq
where $S=\tr  {\bf 1_{\ssc D}}$ is the number of spinor components. This reproduces  
the result in \cite{MS} and \cite{KS}.  

If instead one solves (\ref{Gfull-naive}) for the scalar condensate one reproduces the result 
given in \cite{BBEG}. It 
differs from (\ref{Scond}) by the absence of the factor $\sqrt{1-(1/2d)}$. This is the 
 correction factor for the volume exclusion effect accounted for by the use of IT's  (cf Fig. \ref{lsb4}) and omitted in the derivation of (\ref{Gfull-naive}).

For vector sources, $\bK= k\,\Gamma\cdot n$, where $\Gamma$ stands for either 
$\Gamma_{\ssc V}^\mu$ or $\Gamma_A^\mu$. Noting that in either case one has 
$(\Gamma\cdot n)^{-1} = (\Gamma\cdot n)$, solutions to (\ref{GIfull}) are of the form 
\beq
G_I= g_I(k) \, (\Gamma \cdot n)^{-1} {\bf 1}_{\ssc C} \, ,  \label{GIsoln1}
\eeq 
where now  
\beq
{1\over 2}\sigma_{\ssc \Gamma}\, g_I^2 +kg_I -1 =0 
\label{g2}
\eeq
with 
$\sigma_{\ssc A}= (d-2)(1-(1/2d))$ for the axial vector case, and 
$\sigma_{\ssc V} = - \sigma_{\ssc A}$ for the vector case. 
Then 
\beq
G= g(k) \,(\Gamma\cdot n)^{-1} {\bf 1}_{\ssc C} \label{Gsoln}
\eeq
with, from (\ref{GIsoln1}) and (\ref{Gfull}), $g(k)=[k \pm (d-2)g_I(k)/2]^{-1}$,  where the plus (minus) sign corresponds to axial vector (vector) source. 
Now 
$g_I(0)=\sqrt{2/\sigma_{\ssc \Gamma}}$. Hence, in the axial vector case one gets  
$g(0) = \sqrt{2/(d-2)}\sqrt{1-(1/2d)}$. This then gives an 
axial vector condensate 
\beq
\vev{\bpsi(x)\,i\gamma^5\gamma^\mu \,\psi(x)} = - NS \sqrt{{2\over (d-2)}}\sqrt{1-{1\over 2d}} 
\; n^\mu \,.
\label{AVcond}
\eeq
In the vector case, however, the resulting vev is imaginary. Indeed, it turns complex for small 
source magnitude $k$, which would appear to indicate that no vector condensate 
actually forms. This does not necessarily imply, however, that other condensates induced in 
the presence of a vector source may not persist at vanishing source. 

Again, the same results are obtained from simply solving (\ref{Gfull-naive}) except for 
the omission of the correction factors $\sqrt{1-(1/2d)}$ present in the exact result (\ref{AVcond}). 

Another operator whose condensate is of interest for LSB-induced gravity theories is 
$\bpsi(x) {1\over 2}(\gamma^5\gamma^\mu \sigma_{\kappa\lambda} + \sigma_{\kappa\lambda}\gamma^\mu\gamma^5)\psi(x)$, where $\sigma_{\kappa\lambda}= {i\over 2}[\gamma_\kappa, \gamma_\lambda]$. This condensate is induced in the presence of an axial vector source, and 
indeed survives in the vanishing-source limit: 
\bea
& & \vev{\bpsi(x){1\over 2} (\gamma^5\gamma^\mu \sigma_{\kappa\lambda} + \sigma_{\kappa\lambda}\gamma^\mu\gamma^5)\psi(x)}   \nonumber \\
& = &  -NS \sqrt{2\over (d-2)}\sqrt{1-{1\over 2d}} \left[g^\mu_\kappa n_\lambda - g^\mu_\lambda n_\kappa \right] 
\, . \label{Ccond}
\eea

Other chiral or Lorentz symmetry-breaking condensates that may be induced involve more complicated operators than those considered so far.  They may, in particular, involve lattice nearest-neighbor (continuum derivative) couplings, 
for example $\bpsi(x) \gamma^\mu U_\mu(x)\psi(x+\hmu)$ or  $\bpsi(x)\gamma^\nu U_\mu(x)
\psi(x+\hmu)$. The latter is of particular interest since, in its continuum limit, corresponds to 
the tensor operator $\bpsi(x)\gamma^\nu\partial_\mu\psi(x)$, for which a non-vanishing 
condensate is a natural starting point for a theory of the graviton as a Goldstone boson \cite{KT}.

Consider then the expectation of the (gauge invariant) operator 
 \beq
  O_{\nu\mu}(x) \equiv  \bpsi(x)\gamma_\nu U_\mu(x) \psi(x+\hmu) - 
\bpsi(x)\gamma_\nu U^\dagger_\mu(x-\hmu) \psi(x-\hmu) \label{Tcond}
\eeq
in the strong coupling limit of the theory (\ref{act2}) with an axial vector source: 
\beq
S= \sum_{x,y} \bpsi(x){\bf M}_{x,y}(U)\psi(y) + \sum_x k\,\bpsi(x) (i\gamma^5\gamma^\mu n_\mu) \psi(x)  \, .\label{act5} 
\eeq
The lowest order contribution given by the graph in Fig. \ref{lsb10cA}(a) is easily evaluated:
\bea
\vev{O_{\nu\mu}}^{(0)} & = & - 2 N \left({-i\over 2 k}\right) \left({i\over 2 k}\right) 
\tr_{\ssc D} [\,\gamma_\nu\gamma_\kappa\gamma_5 \gamma_\mu \gamma_\lambda\gamma_5]
n^\kappa n^\lambda \nonumber \\ 
& = & - 2N S \left({1\over 2k}\right)^2 \left[\, 2 n_\nu n_\mu - g_{\nu\mu}\,\right] \, . 
\label{T0}
\eea  
Higher order corrections are obtained by attaching n-trunk tree structures to both ends in Fig. {\ref{lsb10cA}(a)  
as shown in Fig. \ref{lsb10cA}(b). 
Summing these trees as in the previous section leads to the full expectation, represented by the graph 
in Fig. \ref{lsb10a}, in terms of $G(x,x)$ which satisfies (\ref{Gfull}). 
Evaluating the graph and using the solution (\ref{GIsoln1}) -  (\ref{Gsoln}) one obtains 
\bea
\vev{O_{\nu\mu}}& = & {1\over 2} \tr\left[ \gamma_\nu G \gamma_\mu G \right]   \\
        & = & -{1\over 2} NS g(k)^2 \left[ 2n_\nu n_\mu - g_{\nu\mu}\right]  \, , \label{Tfull1}
\eea
which for vanishing source $k=0$ gives : 
\beq
\vev{O_{\nu\mu}}=   -{1\over 2} NS \left({2\over d-2}\right)\left( {1- {1\over 2d}}\right) \left[ 2n_\nu n_\mu - g_{\nu\mu}\right] \, . \label{Tfull2}
\eeq
(\ref{Tfull2}) is a non-vanishing tensorial condensate {\it not} proportional to $g_{\nu\mu}$, i.e. 
a $SO(4)$-breaking (Lorentz-breaking) condensate. (A tensorial condensate proportional to the 
metric tensor is not Lorentz-breaking.) 

It should be noted  that employing vector sources can also induce such a condensate since the attached pair of full tree structures (Fig. \ref{lsb10a}) can be a pair of complex conjugate solutions of the vector version of (\ref{g2}) thus giving a real condensate for all $k$. 

(\ref{Tfull2}) represents partial symmetry breaking. 
The same mechanism, however, can result in any pattern of breaking, partial or complete, by including fermions of different flavors. Different flavors may be coupled to sources 
of different orientation $n_i^\mu$ for each fermion flavor $\psi^i(x)$. 
In physical terms, this may be easily  
envisioned as initial random fluctuations due to some additional flavor-dependent interactions, 
which are much weaker than the strong color forces at the scale (the lattice spacing in our model)  
where the latter drive condensate formation.  
If $N_F$ flavors are present (\ref{Tfull2}) becomes ($d=4$)
\beq
\vev{O_{\nu\mu}}=   {7\over 16} NN_FS \left[ \,g_{\nu\mu} - 
{2 \over N_F} \sum_i   n^i_\nu n^i_\mu \right] \, . \label{Tfull3}
\eeq 
Clearly, any degree of symmetry breaking can be induced in this manner. The strongly coupled 
lattice model considered here provides in fact an explicit realization of the scenario envisioned 
in \cite{KT}. 
\bwt
\
\begin{figure}[ht]
\includegraphics[width=0.7\columnwidth]{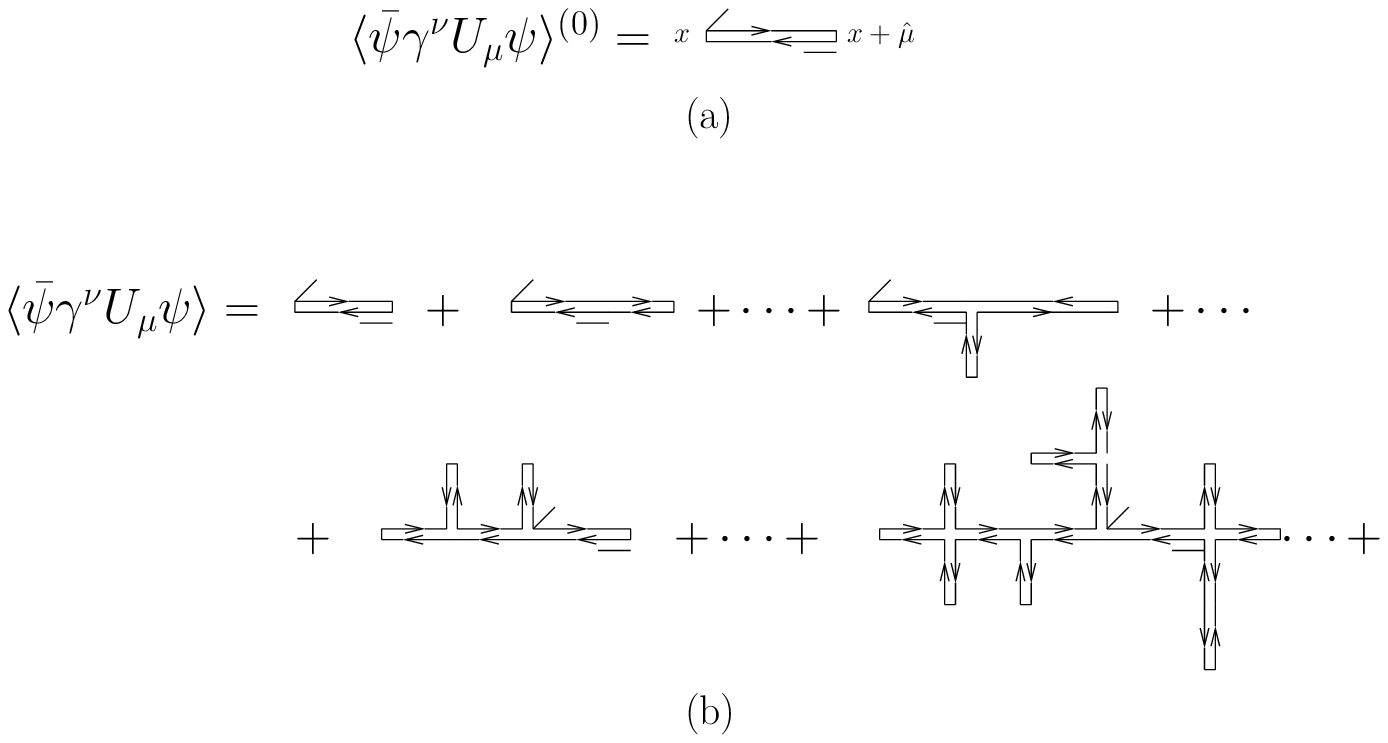} 
\caption{(a) Lowest order graph contribution; (b) sum of graphs for the large-$N$-expectation after attaching all trees at both ends of lowest order graph. The short lines represent the different directions of the $\gamma^\nu$, $\gamma^\mu$ factors in the expectation.  \label{lsb10cA}}
\end{figure}
\ewt

That the tensor condensates (\ref{Tfull2}), (\ref{Tfull3}) were here arrived at by the 
introduction of axial sources is not ultimately relevant in the following sense. When all sources are turned off the surviving condensate gives the actual ground state. Thus, if it corresponds to, say, complete symmetry breaking, 
other Lorentz non-invariant operators, like vectors and axial vectors, may in general have non-zero expectations in this state.  

\begin{figure}[htb]
\includegraphics[width=0.73\columnwidth]{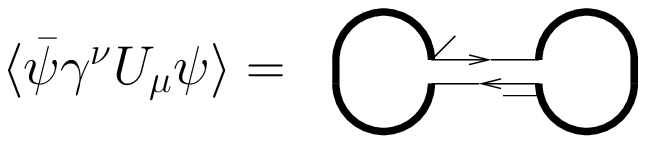} 
\caption{Summation of trees in Fig. \ref{lsb10cA}(b) giving single graph in terms of $G$ for the 
expectation.  \label{lsb10a}}
\end{figure}

One may, for contrast, also consider the effect of the scalar condensate on (\ref{Tcond}). 
Repeating the calculation with a scalar source replacing the axial vector source in (\ref{act5}) one now easily obtains 
\beq
\vev{O_{\nu\mu}}=   {1\over d}\left( {1- {1\over 2d}}\right) NN_FS  \,g_{\nu\mu} \label{Tfull4}
\eeq  
instead of (\ref{Tfull3}). Thus, as expected, no Lorentz symmetry breaking is 
induced by the scalar condensate (\ref{Scond}). 

Let us also comment here on the use of other fermion formulations, in particular staggered fermions. 
The naive fermion degrees of freedom (in $d=4$), i.e. 4-component Dirac spinors on each site and their 16 doublers,  are equivalent in the staggered formulation to four flavors of Dirac fermions each coming in four `tastes'. 
Recall (see e.g. \cite{R}) that this is shown by a `spin diagonalization' transformation $\psi(x) = A_x \chi(x)$ by a 
unitary matrix $A_x$ such that, written in terms of the fields $\chi(x)$, the naive fermion action 
becomes diagonal in the Dirac indices. With the Dirac components decoupled, one can keep 
just a single component per site which gives the standard minimal staggered fermion formulation. 
The single component fields on the $2^d$ vertices of a hypercube are then taken to describe 
the $2^{d/2}$ components of a Dirac field with $d^{d/2}$ flavors. If $k$ fields are retained at 
every site there are $k$ `tastes', and the theory possesses a $U(k)$ or, in the massless case, a 
$U(k)\times U(k)$ global symmetry; $k=4$ corresponds to the $d=4$ naive fermion case. 

Consider then our various condensate operators above such as 
$\bpsi(x)\Gamma^A\psi(x)$. The corresponding operators in the staggered fermion formulation 
are clearly not obtained simply by a spin diagonalization rewriting in terms of the staggered 
fields $\chi(x)$ since the latter are not assigned spinor transformation properties in the same way as naive fermions. One cannot obtain quantities with definite transformation properties 
(spinor, vector, etc) formed only from $\chi$, $\bar{\chi}$ fields at site $x$. Indeed, recall 
(e.g. \cite{R}) that the flavor-spinor content of staggered fermions is displayed by taking appropriate linear combinations of the fields on the $2^d$ vertices of a hypercube, i.e. going to the flavor basis.  If $x_\mu=2y_\mu + \rho_\mu$, with $\rho=0,1$, denote the vertex coordinates of a hypercube labeled by coordinates $y_\mu$ (even length lattice), the block fields 
$q^{a i}_\alpha(y) \equiv \sum_\rho \Gamma^a_{\rho; \alpha} \chi^i(2y+\rho) $ 
are Dirac spinors ($\alpha=1,\cdots, 2^{d/2}$), with $2^{d/2}$ flavors indexed by $a$ and $k$ tastes indexed by $i$.  (An explicit realization of the $\Gamma_\rho$ matrices is given by 
$\Gamma_\rho = \gamma_1^{\rho_1}\gamma_2^{\rho_2}\cdots \gamma_d^{\rho_d}$.) 
In terms of these fields then the condensate operators of interest here are  
\beq 
\bar{q}^{ai}(y) \gamma^\mu q^{ai}(y), \qquad  \bar{q}^{ai}(y)\gamma^5 \gamma^\mu q^{ai}(y) \, , \qquad \mbox{etc} \; . \label{stagconds}
\eeq

It is well-known that though the `local' staggered formulation in terms of the $\chi(x),\bar{\chi}(x)$ fields and the flavor-representation in terms of the `block' $q(y), \bar{q}(y)$ fields are unitarily 
equivalent for free fermions, this is not true in the presence of interactions. Various undesirable features are known to occur \cite{GS} in the flavor (or Dirac-K\"{a}hler) representation of the theory solely in terms of the $q(y), \bar{q}(y)$ fields. 
For this reason it is generally not used. In the present context this means that one would, as usual,  employ the  
`local' form of the staggered action and view the operators of interest (\ref{stagconds}) as defined 
in terms of linear combinations of the $\chi(x),\bar{\chi}(x)$.  
With minimal staggered fermions ($k=1$) the number of flavors due to doublers is reduced by a factor of four. This, however, is of no significance in the present context. On the contrary, extra 
flavors are welcome and can be put to good use as we saw in connection with condensates such as (\ref{Tfull3}) and as we will see in the next section. 
Clearly, there appears to be no particular advantage in using staggered fermions. Naive fermions offer a simpler and more elegant and straightforward framework, and are indeed best suited for our purposes.

\section{Locking of internal and space-time symmetries}\label{L}

When  internal (global) symmetry groups are present, 
the formation of the $SO(4)$ (Lorentz) symmetry-breaking condensates considered so far is not the 
only possibility. A further possibility arises, i.e. condensate formation that `locks'  space-time and internal symmetries. This possibility can be 
equally well explored within our strong coupling lattice gauge models. 

The most straightforward example is provided by taking the internal symmetry to be a copy 
of the (Euclidean) space-time symmetry, i.e. an internal $SO(4)$ group with the fermions transforming as Dirac spinors under it.  Denoting the gamma matrices acting on the internal space by 
$\gamma^m$  and $\hat{\gamma}^5$, consider the  operators 
\beq 
{O_{\ssc V}}^n_\nu = \bpsi(x) \gamma^n\,(i\gamma_5\gamma_\nu)\psi(x) ,  \quad  
{O_{\ssc A}}^n_\nu = \bpsi(x) \hat{\gamma^5}\gamma^n\, \gamma_5\gamma_\nu\psi(x)   
\label{lockop}    
\eeq
involving an external axial vector and an internal vector or axial vector. Non-vanishing vev's of such fermion bilinears can lead to locking between the corresponding groups. 

To compute these vev's we proceed as before. We introduce a source coupled to either of these operators in the action 
so that it is given by (\ref{act4}) but now with the fermions carrying also an internal group index. $G(x,x)$ is a matrix in color, spinor and internal spinor space, and the source $\bar{K}$ is defined as
\beq
\bar{K}=  k ({\Gamma_{\rm I}}^m l_m)\, (i\gamma^5\gamma^\mu n_\mu) 
\, .\label{act6}
\eeq
In (\ref{act6}) the short-hand ${\Gamma_{\rm I}}^m$, ${\rm I}=V,A$, stands for either ${\Gamma_V}^m=
\gamma^m$ or ${\Gamma_A}^m = i\hat{\gamma}^5\gamma^m$, and   
$n^\mu$ and $l^m$ denote arbitrary unit vectors in the external and internal carrier space, respectively.

With these substitutions one has only to repeat our previous computation 
leading to (\ref{GIfull}) and (\ref{Gfull}) with $\bar{K}$ given by (\ref{act6}). So now (\ref{GIfull}) is solved by 
\beq
G_I = g_I(k) \,({\Gamma_{\rm I}}^m l_m)^{-1}(i\gamma_5\gamma_\mu n^\mu)^{-1} 
{\bf 1}_{\ssc C} ,  \label{GILsoln1}
\eeq 
where $g_I(k)$ again satisfies (\ref{g2}) for the axial case, i.e. $g_I(0) = \sqrt{2/\sigma_{\ssc A}}$
with $\sigma_{\ssc A}=(d-2)(1 -(1/2d))$; and 
$G = g(k) \,({\Gamma_{\rm I}}^m l_m)^{-1}(i\gamma_5\gamma_\mu n^\mu)^{-1} 
{\bf 1}_{\ssc C}$ 
is determined by substituting in: 
\beq
G^{-1} = \left[ \, k ({\Gamma_{\rm I}}^m l_m) (i\gamma_5\gamma_\mu n^\mu) {\bf 1}_{\ssc C} + {1\over 2}   \gamma^\mu G_I \gamma_\mu 
\,\right]  \, .\label{EFL}
\eeq 
The vev for (\ref{lockop}) are then given by 
\beq
\vev{{O_I}^n_\nu} = - \tr \left[G \,{\Gamma_{\rm I}}^n (i\gamma_5\gamma_\nu)\right]
\eeq
Taking an internal axial vector one thus obtains 
\beq
\vev{\bpsi(x) \hat{\gamma}^5\gamma^n \, \gamma_5\gamma_\nu\psi(x)} = 
NS^2\, \sqrt{{2\over (d-2)}}
\sqrt{1- {1\over 2d}} \, 
 l^n n_\nu \,. \label{Laa}
\eeq

Different fermion flavors can be coupled to different sources. Take the number of 
flavors to be (a multiple of) four in $d=4$ space-time dimensions. Let $n_{(i)}^\mu$ and $n_{(i)}^m$, $i=1, \ldots, 4$, denote 
a set of orthonormal tetrads in external and internal space, respectively. 
Coupling source
\beq
\bar{K}_{(i)}=  k ({\Gamma_{\rm I}}^m n_{(i)\,m})\, (i\gamma^5\gamma_\mu n_{(i)}^\mu) 
 \label{act7}
\eeq
to the $i$-th (subset of) flavor, and repeating the calculation leading to (\ref{Laa}), one now gets 
\bea
\vev{\bpsi(x)  \hat{\gamma}^5\gamma^n\,\gamma_5\gamma_\nu\psi(x)}&  = & 
\sqrt{{7\over 8}} NS^2 \sum_i  n_{(i)}^n n_{(i)\,\nu} \nonumber \\
& = & \sqrt{{7\over 8}}NS^2 \delta^n_\nu  \,. \label{CLaa}
\eea
(\ref{CLaa}) represents complete locking of the internal and external symmetry, i.e. breaking to the diagonal $SO(4)$ subgroup of the original symmetry: $SO(4) \times 
SO(4)_I \to SO(4)_D$. The condensate remains invariant only under simultaneous equal internal and external rotations. There is, of course, no flavor breaking after the sources are turned off.

It is interesting to note that, in the present context,  the lattice fermion doublers can be taken to 
supply additional flavors. It is indeed amusing to observe that, in terms of the equivalent staggered fermions, the model would automatically possess an $SU(4)\times SU(4)$ chiral invariance, which, 
after formation of the condensate (\ref{Scond}), would break to the diagonal subgroup providing 
precisely four flavor degrees of freedom. 
At any rate, by varying the fermion content models exhibiting partial or complete locking between the external and internal groups can be produced by non-vanishing condensates 
such as (\ref{Laa}). In particular, generalizations can be considered where the internal group $SO(4)_I$ above is replaced by any internal global symmetry group possessing an $SO(4)$ subgroup and an operator transforming as (axial) vector under this subgroup.

This may be all well and good in Euclidean space, where the external group is compactified. 
At this point, however,  the obvious question is: how can such locking work in Minkowski space? 
This issue does not arise in the case of the condensates considered in the previous sections. There 
passage to Minkowski space involves nothing more than the standard Wick rotation, after which 
any condensates such as (\ref{AVcond}), (\ref{Tfull3}) simply transform under the uncompactified external $SO(3,1)$. In the case of locking between the external and an internal group, however, the passage to Minkowski clearly requires more consideration. There appear to be two possible choices. 

One choice is the standard procedure. One makes the usual passage to Minkowski space by the standard Wick rotation. The external group 
gets decompactified to $SO(3,1)$ whereas the internal group remains compact. 
The condensate (\ref{CLaa}) is now invariant only under simultaneous $SO(3)$ (spatial) rotations, i.e. 
$SO(3,1)\times SO(4)_I \to SO(3)_D$. 

The second possibility is to define the passage to Minkowski space to also involve a `Wick rotation' 
of the internal group decompactifying it. Full locking then is preserved, i..e (\ref{CLaa}) remains invariant under $SO(3,1)_D$. This, however, presents an obvious difficulty: one now has an internal non-compact group, such as $SO(3,1)$, which possesses only non-unitary finite-dimensional representations. This, of course, leads in general to unitarity violation. (Unwanted negative signs in propagator residues introduced by the indefinite 
internal group metric is commonly the most direct manifestation of this.) 
The only way out is to take the fermions to transform under a unitary, i.e. an infinite dimensional representation of the internal non-compact group. There appears no problem of principle in doing so. Contrary to the case of external (space-time ) groups, for an internal symmetry the usual formalism applies whether one 
uses finite or infinite dimensional unitary representations \cite{F7}.

\section{Discussion and outlook}\label{D}
We have examined the formation of fermionic condensates in lattice gauge theories in the limits of 
strong coupling and large $N$. We noted that a variety of condensates are directly related to the 
fermion 2-point function at coincident points, $G(x,x)$. A self-consistent equation satisfied by this object was then derived. This was done in two equivalent ways: either by re-summation of the dominant graphs at large $N$ in the fermion hopping expansion at strong coupling; or by direct 
computation of the effective action for composite operators at large $N$ and strong coupling.  
Solutions to the resulting equation for $G(x,x)$ allow then for formation of various 
condensates. Previous results concerning the formation of the chiral 
symmetry breaking condensate in this limit were recovered. Furthermore, certain non-vanishing Lorenz symmetry breaking condensates were obtained. These include the axial vector condensate and the rank-2 tensor condensate. 

The efficacy of axial vector couplings in Lorentz-breaking condensate formation (section 3) is noteworthy. Intimations of the importance of axial coupling were obtained before in the context of 
condensate formation at finite chemical potential (which introduces explicit 
Lorentz symmetry breaking) \cite{Sa}, but also from other more general arguments \cite{AJO}.  
All this may suggest that chiral gauge theories in a strong coupling regime perhaps possess 
the right dynamics for Lorentz-breaking phases. 

Our results on condensate formation were obtained in the large $N$ limit. In the case of chiral symmetry breaking it is of course known, by standard QCD phenomenology and 
ab initio Monte Carlo simulation, that the condensate actually persist for all $N$. 
It is not known at this stage what the fate of the Lorentz-breaking condensates is as $N$ is varied,  
even within the strong-coupling limit.  The large $N$ limit provides a well-defined 
gauge-invariant non-perturbative model that is tractable. The non-leading $1/N$ corrections, however, are much harder to treat. One can expect that the results 
persist for sufficiently large $N$ when the corrections are small. This is important since strictly infinite $N$ may prevent taking the continuum limit by the presence of phase transitions 
(see \cite{Nar} for review). When the corrections become sizable by sufficiently lowering $N$,  
however, the question of the fate of any particular condensate becomes entirely open. 
As we saw, already at infinite $N$ apparently certain condensates form but not others. 
The difference between the vector and axial vector cases, in particular, can be traced to the different 
signs in corresponding contributing graphs. When finite $N$ corrections become important,  
the signs of their contributions will be crucial in determining the stability of a particular condensate. 
In this connection, it should be recalled that effective field theory analysis shows that 
sufficiently large $N$ is crucial for the stability of the low energy effective theory under  
radiative corrections \cite{KT}. 
All this might indicate that a large number of degrees of freedom is necessary for 
formation of stable Lorentz-breaking condensates. 

It would certainly be worthwhile to try to answer these questions by Monte Carlo simulations. 
The strong coupling lattice model provides the simplest case for an initial exploration, even though 
signals will be quite noisy.  Any simulation attempts, however, will be hampered by not 
knowing how large ``sufficiently large" $N$ has to be.

In this paper 
we also found that, within our large $N$ strong coupling lattice models, another type of condensate 
may form, namely condensates locking internal and external symmetries. This possibility appears not to have been considered before in the 
context of dynamical symmetry breaking via condensate formation. 
As discussed in the previous section, 
complete locking with the Lorentz group in Minkowski space requires non-compact internal groups, and this 
necessitates fields transforming under unitary, hence infinite dimensional, 
representations of such internal groups. 
There appears to be no problem in principle with such an assignment. 
At any rate, motivated by the lattice model result, one could postulate the existence 
of a condensate which locks the Lorentz group to an internal non-compact group $G$ containing 
an $SO(3,1)$ subgroup by breaking down to the diagonal $SO(3,1)$ subgroup. 
The resulting low energy theory is then invariant under the 
unbroken diagonal $SO(3,1)$ with the condensate providing a dynamically generated background 
vierbein field connecting internal and external indices. This is an approach to the potential 
construction of a quantum gravity theory that has not be explored before. 
It would certainly be interesting to work out the low
energy effective theory for the Goldstone bosons of the non-linearly realized symmetry
for different choices of the group $G$.

An important issue we have not addressed here is the following. We considered the 
formation of each specific condensate in isolation, induced by the introduction of an appropriate external source which is eventually turned off. We did not consider the interference or competition between different condensates. Formation of a symmetry-breaking condensate 
generally implies the formation of a tower of `higher' ones, both in the chiral and Lorentz 
cases. There may, however, be competition between distinct classes of condensates leaving different subgroups intact, one class resulting in a lower vacuum energy state than another. 
An example could be the formation of complete Lorentz-symmetry-breaking condensates versus 
that of internal-external symmetry-locking condensates that leave a Lorentz subgroup intact. 
Though certainly possible to explore such questions with the techniques used here, 
in particular the composite operator effective action, it requires more involved computations than those carried out here. 
\vspace{1cm}

The author would like to acknowledge the hospitality at the 
Center for Theoretical Physics, MIT, and the Aspen Center for Physics (workshop `Critical behavior of lattice models') where 
most of the work in this paper was performed. 
He would also like to thank P. Kraus, R. Jackiw, A. Niemi, H. Neuberger, V.P. Nair, A. Jenkins, 
M. Ogilvie and F. Sannino for discussions, and J. Greensite for correspondence.   
This work was partially supported by NSF-PHY-0852438. 



\begin{thebibliography}{99}
\bibitem{F0} In the case of QCD, however, which is in one phase for all values of the bare coupling, these features can be expected to persist in the continuum limit.
\bibitem{F1} Closed paths not so coupled give disconnected graphs canceling against the denominator.
\bibitem{F2} Indeed, the rules of Grassmann integration automatically take into account the exclusion principle: 
the path-integral (signed) summation over all fermionic configurations gives the appropriate cancellations 
between configurations contributing to connected and disconnected graphs. As a result, after the factoring of disconnected graphs, one is left with unrestricted sums for the remaining connected 
graphs \cite{R}.
\bibitem{F4} Another similar example \cite{CJT} is the large $N$ limit of the $O(N)$ 
model, where again a single 2-PI graph replaces the direct summation \cite{CJP} of the infinite set of leading $N$ graphs .
\bibitem{F5} Picking out particular choices of pairs in products of Grassmann variables inside a Grassmann  integral to be replaced by bosonic variables is tricky. 
Any approximations generally render the procedure ambiguous. In the 
present case note that the powers of $\bar{A} A$ can be re-expressed as powers of 
different pairings of Grassmann fields by Fierz rearrangements. In this manner one may equally well 
introduce other composite fields that may be similarly replaced by bosonic variables. 
Any subsequent approximations lead to the ambiguities of what becomes essentially an ad-hoc Hartree-Fock type of approximation. Appeal to special factorization properties at large $N$ may 
in certain cases justify such manipulations.
\end{thebibliography}
\end{document}